\documentclass[aps,pra,letterpaper,10pt,twocolumn]{revtex4-1}
\usepackage[colorlinks=true,allcolors=blue]{hyperref}
\usepackage{amssymb}
\usepackage{amsmath}
\usepackage{graphicx}
\usepackage[caption=false]{subfig}
\usepackage{amsfonts}
\usepackage{xcolor}
\usepackage{epsfig}
\usepackage{color}
\usepackage{bm}
\usepackage{tabularx}
\usepackage{multirow}
\usepackage{mathtools}
\usepackage{xfrac}
\usepackage{blkarray}
\usepackage{bbold}
\usepackage[mathscr]{euscript}
\usepackage{autobreak}
\usepackage{comment}

\newcommand{\cncb}{C$_{\rm N}$C$_{\rm B}$}
\newcommand{\vn}{V$_{\rm N}$}

\newcommand{\cn}{C$_{\rm N}$}
\newcommand{\cb}{C$_{\rm B}$}

\begin{document}
\title{Ultraviolet quantum emitters in $h$-BN from carbon clusters}

\author{Song Li}
\affiliation{Wigner Research Centre for Physics, P.O.\ Box 49, H-1525 Budapest, Hungary}

\author{Anton Pershin}
\affiliation{Wigner Research Centre for Physics, P.O.\ Box 49, H-1525 Budapest, Hungary}

\author{Gerg\H{o} Thiering}
\affiliation{Wigner Research Centre for Physics, P.O.\ Box 49, H-1525 Budapest, Hungary}

\author{P\'eter Udvarhelyi}
\affiliation{Wigner Research Centre for Physics, P.O.\ Box 49, H-1525 Budapest, Hungary}

\author{Adam Gali}
\thanks{gali.adam@wigner.hu}
\affiliation{Wigner Research Centre for Physics, P.O.\ Box 49, H-1525 Budapest, Hungary}
\affiliation{Budapest University of Technology and Economics,  Budafoki \'ut 8, H-1111 Budapest, Hungary}

\date{\today}
\begin{abstract}
Ultraviolet (UV) quantum emitters in hexagonal boron nitride (hBN) have generated considerable interest due to their outstanding optical response. Recent experiments have identified a carbon impurity as a possible source of UV single photon emission. Here, based on the first principles calculations, we systematically evaluate the ability of substitutional carbon defects to develop the UV colour centres in hBN. Of seventeen defect configurations under consideration, we particularly emphasize the carbon ring defect (6C), for which the calculated zero-phonon line (ZPL) agrees well the experimental 4.1-eV emission signal. We also compare the optical properties of 6C with those of other relevant defects, thereby outlining the key differences in the emission mechanism. Our findings provide new insights about the large response from this colour centre to external perturbations and pave the way to a robust identification of the particular carbon substitutional defects by spectroscopic methods.

\end{abstract}

\maketitle

%
%
\section{Introduction}
Single point defects in two-dimensional (2D) hexagonal boron nitride (hBN) play a vital role in the optical properties of the host and hold great promise for quantum information technologies and integrated quantum nanophotonics~\cite{tran2016quantum, gottscholl2020initialization, chejanovsky2021single, mendelson2021identifying, hayee2020revealing, bourrellier2016bright, bommer2019new, tran2016robust}. As compared to the bulk counterparts, the reduced dimensionality and spatial confinement of wavefunctions enable a more viable integration of 2D hBN with external materials to form the quantum architectures. In particular, colour centres in hBN are responsible for ultrabright single-photon emission at room temperature with a wide range of emission wavelengths~\cite{tran2016robust, sajid2020single}. Recent experiments demonstrated the versatile properties of the defect emitters in 2D hBN, such as a strain and electric field dependent emission~\cite{grosso2017tunable, hayee2020revealing, mendelson2020strain, noh2018stark}, high stability under high pressure and temperature~\cite{xue2018anomalous, kianinia2017robust, vokhmintsev2021temperature}, as well as initialization and readout of a spin state through optical pumping~\cite{gottscholl2020initialization, gottscholl2021room}. Other studies have shown a successful engineering and coherent control of a single spin in hBN~\cite{chejanovsky2021single}, whilst the room temperature initialization and readout have also been realized~\cite{gottscholl2020initialization, gottscholl2021room}.

Of several photoluminescence (PL) signals from the colour centres in hBN, a strong ultraviolet (UV) emission at close to $\sim$4.1~eV has received much of attention~\cite{museur2008defect, watanabe2004direct, du2015origin, vuong2016phonon, pelini2019shallow}. Noteworthy, the deep-UV emission permits the optical operations under the daily light due to a vanishing overlap with the solar radiation spectrum. The single photon emission associated with these bands indicates that it should originate from a point defect~\cite{bourrellier2016bright, tan2019ultraviolet}. However, despite various attempts, the atomistic origin of the UV emission in hBN is still under debate. In particular, due to the similarities with the carbon-doped hBN samples (mostly due to the PL lifetime of $\sim$1.1~ns~\cite{museur2008defect, era1981fast}), carbon is thought to contribute into the formation of the PL signal~\cite{uddin2017probing, du2015origin}. Many theoretical attempts, mainly based on the density functional theory (DFT) calculations, have outlined several possible defect configurations for the 4.1-eV emission. More specifically, an earlier study indicated that the recombination from a donor-acceptor pair (DAP) involving \cn\ and \vn\ \cite{du2015origin} might be related to the 4.1-eV emission. However, Weston \textit{et al.} argued that the donor level of \vn\ is deep in the gap~\cite{weston2018native} while the spatial separation between the two is unlikely to explain the short PL lifetime. Instead, they proposed the \cb\ as a possible source, owing to the charge transition level (CTL) $(0/+)$ at 3.71~eV~\cite{weston2018native}. Furthermore, Mackoit \textit{et al.} studied the carbon dimer \cncb\ of which calculated zero-phonon line (ZPL) at 4.3~eV, the calculated optical lifetime and Debye-Waller factor could well explain the optical properties of 4.1-eV emitters~\cite{mackoit2019carbon}. Other more complex carbon-related defects, involving up to ten carbon atoms, were also investigated~\cite{korona2019exploring, jara2021first}. On the other hand, some 4.1-eV UV emitters did not show carbon-related isotope shift in the phonon sideband~\cite{pelini2019shallow}, and an intrinsic Stone-Wales defect was proposed as the origin for those UV emitters, with obtaining ZPL at 4.09~eV~\cite{hamdi2020stone}. Despite some of the proposed configurations exhibit the excitation energies at around 4~eV, many of their key properties, including the stability, electronic configuration, and vibronic properties were not considered. Recently, additional lines were observed in the 4.1--4.2~eV range and the isotopically controlled carbon doping is employed to determine the role of carbon impurity to the 4.1-eV emission~\cite{pelini2019shallow}. In particular, the additional lines, distinct from the previous 4.1-eV emission, show strong PL intensity with a clear temperature-dependency~\cite{vokhmintsev2021temperature}. These findings motivated us to carry out a systematic theoretical study to reveal the role of substitutional carbon defects in the formation of the UV single-photon emitters in hBN.

In this paper, we analyse seventeen configurations of substitutional carbon defects and systematically address their thermodynamic properties. Among those, we identify a six carbon ring defect, where the carbon atoms substitute one BN honeycomb of hBN lattice, as one stable defect configuration. Noteworthy, this defect have been already unambiguously identified through the annular dark field scanning transmission electron microscopy (ADF-STEM)~\cite{krivanek2010atom, park2021atomically}, and can be intentionally introduced into the lattice with atomic precision by the focused electron beam~\cite{park2021atomically}. We show that this colour centre emits light due to a strong electron-coupling with $E$-phonon modes, caused by the product Jahn-Teller effect. More specifically, the respective symmetry lowering is found to activate a forbidden transition through an intensity borrowing mechanism from a higher-lying bright state. We further calculate the ZPL energy, luminescence spectrum, and radiative lifetime and found them in excellent agreement with the experimental observations for the 4.1-eV emission. In addition, we examine the isotopic shift in ZPL and sideband, caused by the presence of ${}^{13}$C isotopes, and compare the results to those of other carbon defect configurations.

\begin{figure*}[tb]
\includegraphics[width=2\columnwidth]{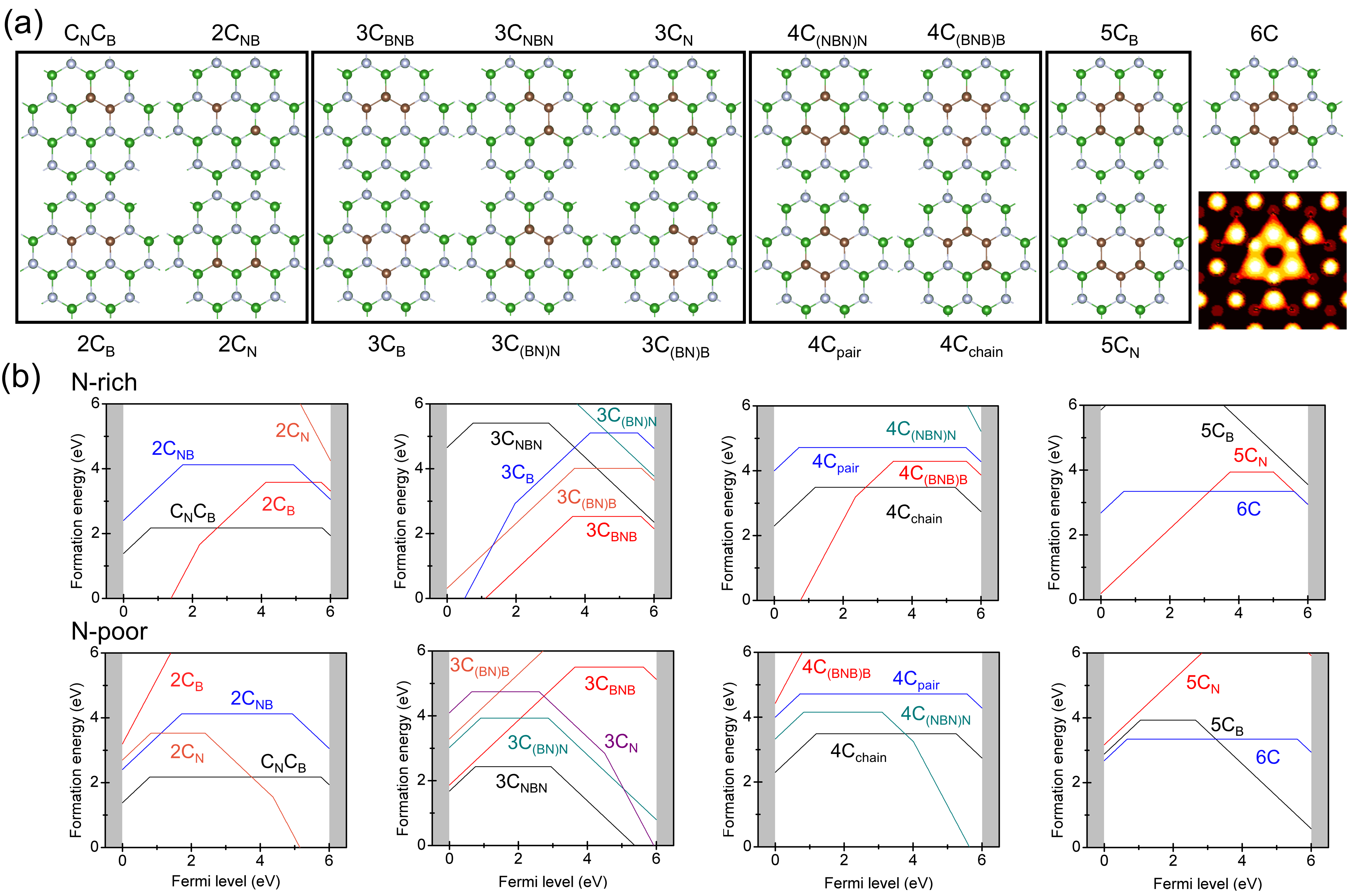}
\caption{\label{fig:1}%
(a) Different carbon defects we considered here and the simulated scanning tunneling microscopy image for 6C defect. (b) Calculated formation energy vs.\ Fermi level under N-rich and N-poor conditions. The grey colour depicts band edge.}
\end{figure*}

\begin{figure*}[tb]
\includegraphics[width=2\columnwidth]{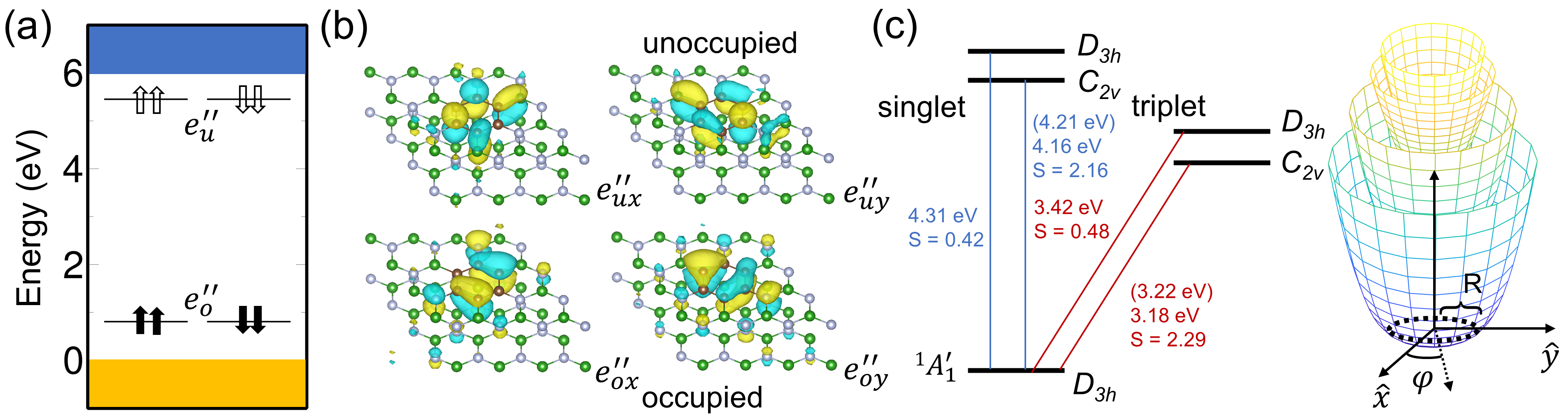}
\caption{\label{fig:2}%
(a) Single particle energy level of carbon ring defect in the ground state. The subscript $o$ and $u$ indicate the occupied and unoccupied defect states while the arrows denote the spin directions. (b) The wave function isosurface of defect levels. (c) The energy diagram of optical transition with zero-phonon line (ZPL) and Huang-Rhys (HR) factor calculated with density funcational theory. The values in parentheses are corrected ZPL with product Jahn-Teller (pJT) effect. Left schematic figure represents the four layers APES of pJT effect. The dash line is the energetically global minima loop.}
\end{figure*}

\section{Results}
\label{sec:results}
To begin with, we systematically analyse the thermodynamic properties of the carbon defects in hBN, with the aim of identifying the possible sources of the UV emission. Since the experimental PL signal features a short radiative lifetime, we only focus on those arrangements where the carbon atoms are closely packed within a single honeycomb. Noteworthy, the delocalization of defect orbitals should naturally decrease the excitation energy; therefore, larger defect complexes were not considered. The resulting structures of seventeen distinct C-configurations are shown in 
Fig.~\ref{fig:1}(a) and Supplementary Figure 1. For those, we evaluated the formation energy diagrams and charge transition levels (CTLs), that are plotted in Fig.~\ref{fig:1}(b). Our calculations confirm a high affinity of hBN towards the formation of substitutional carbon defects, since for most of them, the formation energy is within 5~eV. Besides, the formation energies for the defects with an unequal amount of substituted B and N can be largely decreased by selecting the appropriate growth conditions. However, for a given number of carbon atoms, we always observe that the most stable configurations represent the confined C-clusters, where the carbon atoms are arranged in a continuous chain. Importantly, to prevent a photoionization process, a UV quantum emitter should maintain a stable charge state. This condition is observed for the defects with an even number of carbon atoms (namely, \cncb, 2C$_\text{NB}$, 4C$_\text{chain}$, 4C$_\text{pair}$ and 6C); they posses a highly-stable neutral charge state over the energy range, exceeding the ionization threshold. By contrast, the defects with an odd number of carbons rapidly change their charge states across the formation energy diagrams owing to their radical nature. Our calculations provide a low formation energy of 2.17~eV the carbon dimer \cncb\ which agrees well with the previous reports~\cite{mackoit2019carbon, maciaszek2021thermodynamics}. In addition, the formation energy of 6C ring is found 1.2~eV larger than that for the dimer (0.5~eV with PBE~\cite{perdew1996generalized} functionals) and this is the second lowest formation energy.

Having identified the 6C ring defect as a stable defect configuration, we now focus on its structural and electronic properties. In the neutral charge state, the ground state configuration of the defect embedded in the hBN layer is a closed-shell singlet, and it exhibits $D_{3h}$ symmetry. The bond lengths between carbon atoms and the nearest neighbour atoms are 1.42~\AA\ and 1.51~\AA\ for C-N and C-B, respectively. The electronic structure of hBN with the 6C ring defect is shown in Fig.~\ref{fig:2}; it features two pairs of degenerate $e^{\prime\prime}$ orbitals where two $e^{\prime\prime}$ orbitals fall close to the valence band maximum, fully occupied by four electrons, and the other two fall close to the conduction band minimum. Of note, this electronic configuration resembles the occupation of the $\pi$ bonding and $\pi^{\star}$ antibonding orbitals of benzene ~\cite{casanova2010quantifying}. In terms of orbital occupation, the electronic configuration reads as $|e^{\prime\prime}_{ox}e^{\prime\prime}_{oy}e^{\prime\prime}_{ux}e^{\prime\prime}_{uy}\rangle$ where $o$ and $u$ indicate the occupied and unoccupied states. This leads to the $^1A^{\prime}_{1}$ symmetry of the ground state. 

From the group theory analysis, the electronic transitions between the $e$ orbitals give rise to four excited states in both singlet and triplet manifolds, expressed as follows: 
\[
{^2}E^{\prime\prime}\otimes {^2}E^{\prime\prime}={^1}A^{\prime}_{1}\oplus {^1}A^{\prime}_{2}\oplus {^1}E^{\prime}\oplus {^3}A^{\prime}_{1}\oplus {^3}A^{\prime}_{2}\oplus {^3}E^{\prime}.
\]
Due to the high degeneracy of the defect orbitals in $D_{3h}$ symmetry, each excited state represents a combination of two Slater-determinants. More precisely, in terms of the single-electron transitions (see Supplementary Note 1), those are given as: 
\[
\left.\begin{array}{c}
|^1A^{\prime}_{1}\rangle=\mathcal{S}\frac{1}{\sqrt{2}}\left(|e^{\prime\prime}_{ox}e^{\prime\prime}_{ux}\rangle+|e^{\prime\prime}_{oy}e^{\prime\prime}_{uy}\rangle\right)\\
|^1E^{\prime}_{x}\rangle=\mathcal{S}\frac{1}{\sqrt{2}}\left(|e^{\prime\prime}_{ox}e^{\prime\prime}_{ux}\rangle-|e^{\prime\prime}_{oy}e^{\prime\prime}_{uy}\rangle\right)\\
|^1E^{\prime}_{y}\rangle=\mathcal{S}\frac{1}{\sqrt{2}}\left(|e^{\prime\prime}_{ox}e^{\prime\prime}_{uy}\rangle+|e^{\prime\prime}_{oy}e^{\prime\prime}_{ux}\rangle\right)\\
|^1A^{\prime}_{2}\rangle=\mathcal{S}\frac{1}{\sqrt{2}}\left(|e^{\prime\prime}_{ox}e^{\prime\prime}_{uy}\rangle-|e^{\prime\prime}_{oy}e^{\prime\prime}_{ux}\rangle\right)
\end{array} \right\}\otimes\mathcal{A}|\uparrow\downarrow\rangle
\]

\[
\left.\begin{array}{c}
|^3A^{\prime}_{1}\rangle=\mathcal{A}\frac{1}{\sqrt{2}}\left(|e^{\prime\prime}_{ox}e^{\prime\prime}_{ux}\rangle+|e^{\prime\prime}_{oy}e^{\prime\prime}_{uy}\rangle\right)\\
|^3E^{\prime}_{x}\rangle=\mathcal{A}\frac{1}{\sqrt{2}}\left(|e^{\prime\prime}_{ox}e^{\prime\prime}_{ux}\rangle-|e^{\prime\prime}_{oy}e^{\prime\prime}_{uy}\rangle\right)\\
|^3E^{\prime}_{y}\rangle=\mathcal{A}\frac{1}{\sqrt{2}}\left(|e^{\prime\prime}_{ox}e^{\prime\prime}_{uy}\rangle+|e^{\prime\prime}_{oy}e^{\prime\prime}_{ux}\rangle\right)\\
|^3A^{\prime}_{2}\rangle=\mathcal{A}\frac{1}{\sqrt{2}}\left(|e^{\prime\prime}_{ox}e^{\prime\prime}_{uy}\rangle-|e^{\prime\prime}_{oy}e^{\prime\prime}_{ux}\rangle\right)
\end{array}\right\}\otimes
\left\{\begin{array}{c}
|\uparrow\uparrow\rangle\\
\mathcal{S}|\uparrow\downarrow\rangle\\
|\downarrow\downarrow\rangle
\end{array}\right.
\]
where the first right-hand side term refers to the orbital part and the second one is for spin part (the arrows indicate the spin directions). Here, we use the antisymmetrization operator, $\mathcal{A}|xy\rangle=\frac{1}{\sqrt{2}}\left(|xy\rangle-|yx\rangle\right)$ for the singlet wavefunctions, and the symmetrization operator, $\mathcal{S}|xy\rangle=\frac{1}{\sqrt{2}}\left(|xy\rangle+|yx\rangle\right)$, for the triplets. Furthermore, each of the single-electron transitions leads to the Jahn-Teller instability for both occupied and empty defect orbitals; this is achieved via a coupling to a quasi-localized $E$ vibration mode and is known as a product Jahn-Teller (pJT) effect~\cite{thiering2019eg, ciccarino2020strong, qiu2007studies}. Thus, the total Hamiltonian, which accounts for both the electronic correlation and pJT, is given as:
\begin{equation}
\begin{split}
\label{eq:tot}
\hat{H}_\text{tot}= \hbar\omega_{E}(a_{x}^{\dagger}a_{x}+a_{y}^{\dagger}a_{y}+1)+\hat{W}+\hat{H}_\text{JT}\text{,}
\end{split}
\end{equation}
where $a_{x,y}$ , $a_{x,y}^{\dagger}$ are ladder operators for creating or annihilating $E$ phonon mode in the two-dimensional space while the first term is the vibrational potential energy of the system. $\hat{W}$ is the electronic Hamiltonian and $\hat{H}_\text{JT}$ is the JT part. 

\begin{figure*}
\includegraphics[width=2\columnwidth]{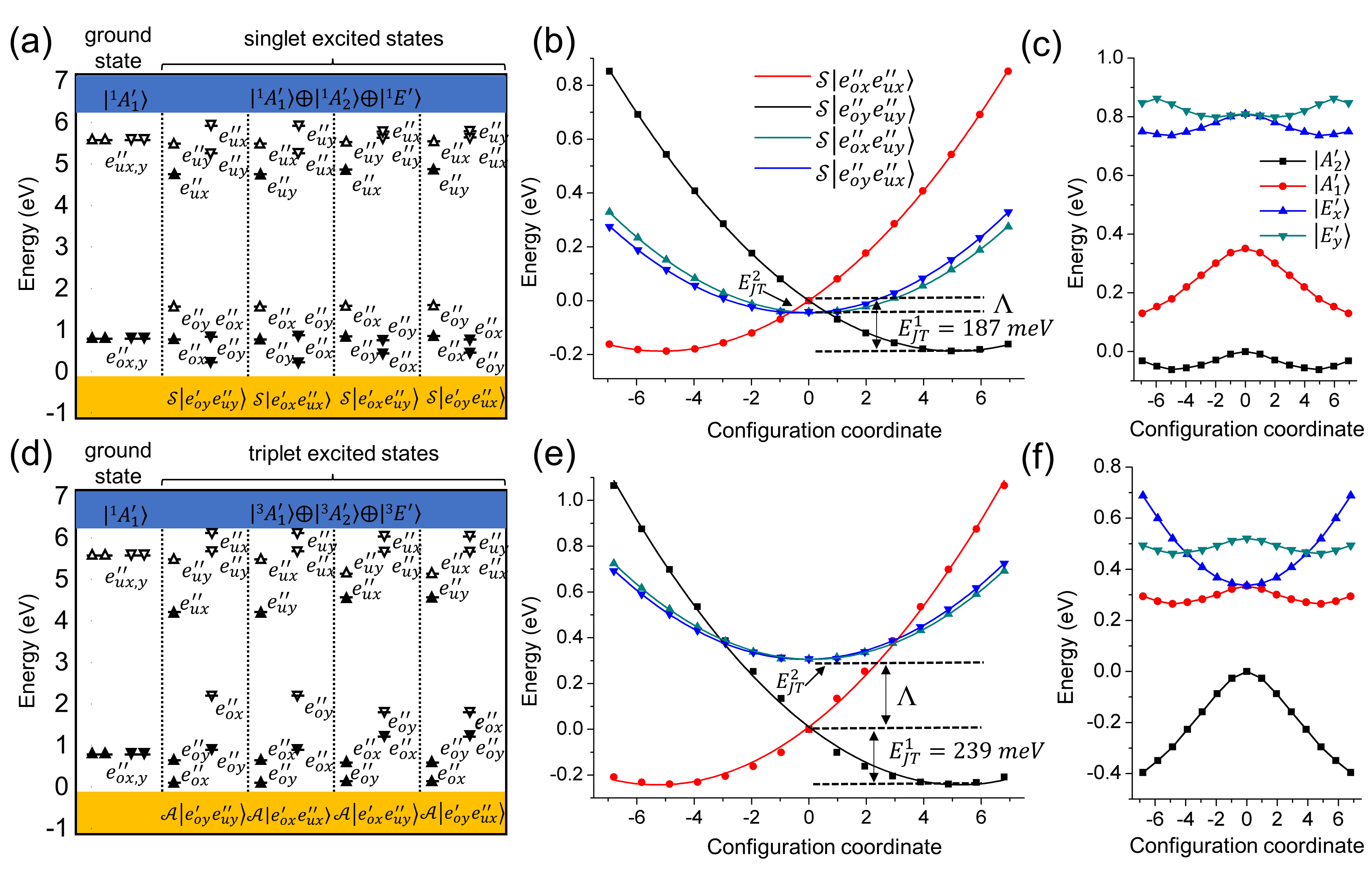}
\caption{\label{fig:3}%
A single-particle energy level diagram of carbon ring defect for singlet (a) and triplet (d) excited states. The filled and hollow arrows indicate the occupied and empty states with up and down spin directions. (b), (d) The calculated APES for singlet and triplet states, respectively. The dots are from DFT result and the solid line is fitted based on pJT model. The standard deviation is less than 3\%. $X = 0$ is the geometry with $D_{3h}$ symmetry and the energy minima could be achieved by removing the symmetry restriction. (c), (e) The energy diagrams for the four states with TDDFT method for singlet and triplet states. The coordinates are built on DFT optimization. The pJT effect is not included here.}
\end{figure*}

In order to solve the $\hat{H}_\text{tot}$, we first construct the $\hat{W}$. Here, the single determinants, which constitute of the wave functions in Eq.~\ref{eq:tot}, are shown in Fig.~\ref{fig:3}(a) and (d). In $D_{3h}$ symmetry, the four single determinants form two double-degenerate branches with $E_d(|e^{\prime\prime}_{ox}e^{\prime\prime}_{ux}\rangle)=E_d(|e^{\prime\prime}_{oy}e^{\prime\prime}_{uy}\rangle)$ and $E_d(|e^{\prime\prime}_{ox}e^{\prime\prime}_{uy}\rangle)=E_d(|e^{\prime\prime}_{oy}e^{\prime\prime}_{ux}\rangle)$, where $E_d$ is the total energy of the (diabatic) state. In the singlet manifold,  $e^{\prime\prime}_{ox}\rightarrow e^{\prime\prime}_{ux}$ (or $e^{\prime\prime}_{oy}\rightarrow e^{\prime\prime}_{uy}$) configurations are stabilized over 41~meV by the exchange interaction (so that $E_d(|e^{\prime\prime}_{ox}e^{\prime\prime}_{uy}\rangle)$ is lower than $E_d(|e^{\prime\prime}_{ox}e^{\prime\prime}_{ux}\rangle)$), while their order is reversed for the triplets. As shown in Fig.~\ref{fig:3}(b) and (e), the energy difference between the two configurations, computed by $\Delta$SCF, are 41~meV and 308~meV for singlets and triplets, respectively.

To provide a robust description of the excited states, we further compute the excitation energies of 6C defect by the second-order approximate coupled cluster singles and doubles model (CC2), thereby focusing on a representative flake model. These calculations were assisted by the time-dependent (TD) DFT to access the transition properties, as well as by two other post-Hartree fock metons (SOS-ADC2 and NEVPT2) for the sake of reference. The resulting (vertical) excitation energies, obtained at the HSE geometry (see Methods), are summarized in Supplementary Table 2. Here, we found that all the approaches consistently predict the appearance of the localized excited states in the energy range between 4 and 5~eV. Of note, at the high symmetry point, the two lowest $A^{\prime}_{1}$ and $A^{\prime}_{2}$ states are dark, while the transitions to $E^{\prime}$ are optically allowed, as evident by the value of oscillator strength of $\sim$0.93 atomic unit. Furthermore, using the definition from Refs.~\citenum{thiering2019eg, ciccarino2020strong}, the electronic Hamiltonian is expressed as follows (see Supplementary Note 2),
\begin{equation}
\hat{W} = \Lambda(|A^{\prime}_{1}\rangle\langle A^{\prime}_{1}|-|A^{\prime}_{2}\rangle\langle A^{\prime}_{2}|)-\Delta(|E^{\prime}_{x}\rangle\langle E^{\prime}_{x}|+|E^{\prime}_{y}\rangle\langle E^{\prime}_{y}|),
\end{equation}
the $A^{\prime}_{1}$ and $A^{\prime}_{2}$ are non-degenerate states and the $E^{\prime}$ is a double degenerate state. The coupling parameters $\Lambda$ and $\Delta$ are then directly read from the CC2 results. Here, we have computed $\Lambda$ and $\Delta$ of $-168.5$ and $-619.5$~meV for the singlets and of $393$~meV and $7$~meV for the triplets, respectively. For the sake of reference, the respective values obtained by TDDFT are $-175.5$~meV and $-634.5$~meV for the singlets, as well as $260.5$~meV and $74.5$~meV for the triplets.

Having defined the $\hat{W}$, we now focus on the pJT Hamiltonian, given as
\begin{equation}
\begin{split}
\label{eq:jte}
\hat{H}_\text{JT}= &F_{o}\left(\hat{\sigma}_{z}\otimes \hat{\sigma}_{0}\hat{x}+\hat{\sigma}_{x}\otimes \hat{\sigma}_{0}\hat{y}\right)\\
&+F_{u}\left(\hat{\sigma}_{0}\otimes \hat{\sigma}_{z}\hat{x}+\hat{\sigma}_{0}\otimes \hat{\sigma}_{x}\hat{y}\right)\text{,}
\end{split}
\end{equation}
where $\hat{\sigma}_{z}=|e_{x}\rangle\langle e_{x}|-|e_{y}\rangle\langle e_{y}|$ and $\hat{\sigma}_{x}=|e_{x}\rangle\langle e_{y}|+|e_{y}\rangle\langle e_{x}|$ are Pauli matrices; $\hat{\sigma}_{0}$ is the unit matrix and $\hat{\sigma}_{0}=|e_{x}\rangle\langle e_{x}|+|e_{y}\rangle\langle e_{y}|$. The major effect of the strong electron-phonon coupling is to drive the excited states out of $D_{3h}$ symmetry to a lower $C_{2v}$. The JT energies, denoted as $E^1_\text{JT}$ and $E^2_\text{JT}$ for $|e^{\prime\prime}_{ox}e^{\prime\prime}_{ux}\rangle$ and $|e^{\prime\prime}_{ox}e^{\prime\prime}_{uy}\rangle$, respectively, are determined by fitting the adiabatic potential energy surfaces (APES) from \textit{ab initio} results, as shown in Fig.~\ref{fig:3}. We found that the JT effect is much more significant for $|e^{\prime\prime}_{ox}e^{\prime\prime}_{ux}\rangle$ than $|e^{\prime\prime}_{ox}e^{\prime\prime}_{uy}\rangle$, which yields the negligible $E^2_\text{JT}$. More specifically, the values of $E^1_\text{JT}$ are $187$~meV and $239$~meV for the singlets and triplets, respectively, while $E^2_\text{JT}$ are only $0.46$~meV and $0.14$~meV. The effective vibration energy $\hbar\omega_{E}$ is then deduced from the lowest branch of APES parabola in a dimensionless generalized coordinates.  Based on these data, the electron-phonon coupling parameters are calculated as,
\begin{equation}
E^1_\text{JT} = \frac{(F_{o}+F_{u})^2}{2\hbar\omega_{E}}, E^2_\text{JT} = \frac{(F_{o}-F_{u})^2}{2\hbar\omega_{E}}\text{.}
\end{equation}
In turn, the linear vibronic Hamiltonian for the last two terms in Eq.~\ref{eq:jte} is given as
\begin{align}
\underbrace{
\begin{array}{lc}
\setlength{\arraycolsep}{0.1pt}
\begin{bmatrix}
\hat{x}(F_{o}+F_{u})&\hat{y}F_{o}&\hat{y}F_{u}&0\\
\hat{y}F_{o}&-\hat{x}(F_{o}-F_{u})&0&\hat{y}F_{u}\\
\hat{y}F_{u}&0&\hat{x}(F_{o}-F_{u})&\hat{y}F_{o}\\
0&\hat{y}F_{u}&\hat{y}F_{o}&-\hat{x}(F_{o}+F_{u})
\end{bmatrix}
\end{array}
}_{|e^{\prime\prime}_{ox}e^{\prime\prime}_{ux}\rangle~~~~~~~~~ |e^{\prime\prime}_{ox}e^{\prime\prime}_{uy}\rangle~~~~~~~~~ |e^{\prime\prime}_{oy}e^{\prime\prime}_{ux}\rangle~~~~~~~~~ |e^{\prime\prime}_{oy}e^{\prime\prime}_{uy}\rangle} \text{,} 
    \label{equation: 44pjt}
\end{align}
where the diagonal part of this expression indicates that with $\pm\hat{x}$ displacement, the energy of single determinants change their energies with constructive and destructive joint vibronic coupling strength $F_{o}~\pm ~F_{u}$; $\hat{H}_\text{JT}$  is a iso-stationary function for the APES of the JT system. 

\begin{figure*}
\includegraphics[width=2\columnwidth]{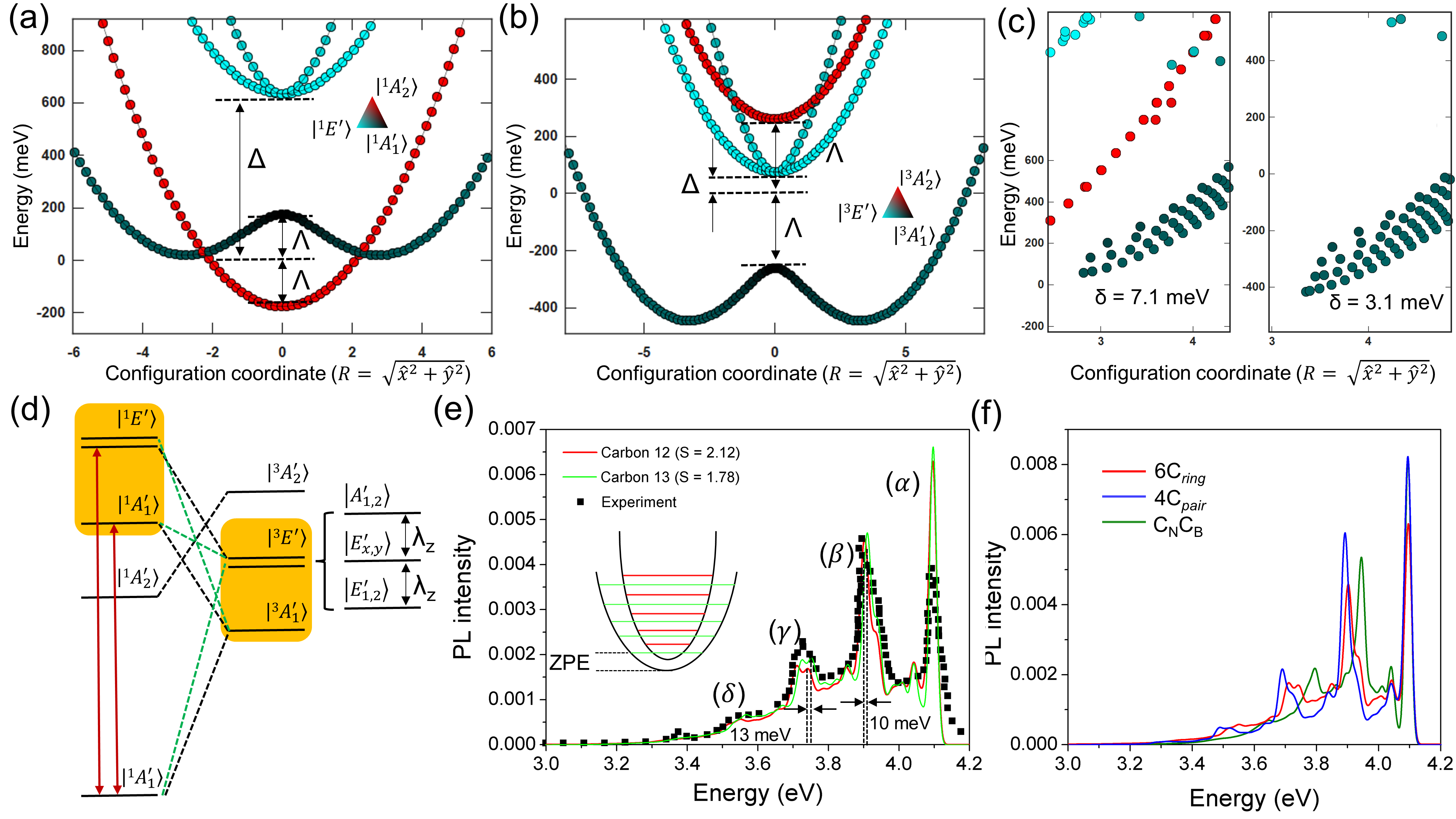}
\caption{\label{fig:4}%
The eigenvalues for the total Hamiltonian of the system in one dimension ($Y = 0$) for (a) singlet and (b) triplet. The pure states of $A^{\prime}_{1}$, $A^{\prime}_{2}$, and $E^{\prime}$ are coloured with black, red and cyan dots. The lowest APES branch is a mixed state of $A^{\prime}_{1}$ and $E^{\prime}$. (c) The polaronic eigenstates for (left) singlet and (right) triplet with full rotation for Eq.~\ref{eq:8}. The second order pJT strength could be estimated by the energy splitting between the two lowest eigenvalues. (d) The schematic energy diagram of electronic states and possible ISC transitions. Black dash line links states with same representation in different spin manifold. Green line links states enabled by pJT induced mixture which happens between states labelled with orange. (e) The simulated PL spectrum (red) and experimental data (black dots). The PSB of isotope $^{13}$C is also shown. The ZPL position is aligned by 0.08~eV to match the first peak in the PSB. The Gaussian broadening is 10~meV. Four peaks can be identified at 4.095~eV, 3.905~eV, 3.711~eV, and 3.551~eV which are consistent with experimental observation. The inner picture is the schematic coordinate diagram of isotopic effect. (f) The simulated PL spectrum of dimer (\cncb), 4C$_\text{pair}$, and 6C ring where the ZPL energies are aligned for the sake of comparison of PSBs.}
\end{figure*}

The solutions for the total Hamiltonian from Eq.~\ref{eq:tot} that incorporate both the vibrational and electronic parts for the singlet and triplet states are plotted in Fig.~\ref{fig:4}. For the singlets in $D_{3h}$ symmetry, the states appear in the following order: E($A^{\prime}_{2}$) $\textless$ E($A^{\prime}_{1}$) $\textless$ E($E^{\prime}$). $A^{\prime}_{2}$ shows no sign of the JT instability or mixture with $E^{\prime}$; thus, it maintains a high symmetry configuration and remains dark along the configuration coordinate. By contrast, when the system is driven out of $D_{3h}$ symmetry, the mixing between the $A^{\prime}_{1}$ and $E^{\prime}$ is clearly apparent. 
In the double degenerate JT system, the vibronic ground states in each $E$ branches is written as
\begin{equation}
|\tilde{E}\rangle = e^{\pm i\varphi}|\tilde{\Psi}_{\pm R,\varphi}\rangle = e^{\pm i\varphi}[\sin\frac{\varphi}{2}|E_x\rangle - \cos\frac{\varphi}{2}|E_y\rangle]\text{,}
\end{equation}
the $e^{\pm i\varphi}$ is a phase factor introduced for the reason that the wave function changes sign when rotating along the bottom of the APES~\cite{Bersuker2006} as indicated by dash line in Fig.~\ref{fig:2}(c). The combination of $|\tilde{E}^{\prime\prime}_{o}\rangle$ and $|\tilde{E}^{\prime\prime}_{u}\rangle$ is 
\begin{equation}
\begin{split}
&|\tilde{E}^{\prime\prime}_{o}\rangle \otimes|\tilde{E}^{\prime\prime}_{u}\rangle = \\
&\frac{1}{\sqrt{2}}
\left\{\begin{array}{c}
1\\
e^{- 2i\varphi}\\
e^{+ 2i\varphi}
\end{array}\right\}[|A^{\prime}_{1}\rangle \pm \cos(\varphi)|E^{\prime}_{x}\rangle \mp \sin(\varphi)|E^{\prime}_{y}\rangle]\text{.}
\end{split}
\end{equation}
This indicates that the minima loop solely constitutes of mixed states of $|A^{\prime}_{1}\rangle$ and $|E^{\prime}\rangle$. The polaronic wave function with this minima loop with full rotation included can be solved by
\begin{equation}\label{eq:8}
\begin{split}
&|\tilde{\Phi}\rangle = \sum_{n,m}[a_{n,m}|e^{\prime\prime}_{ox}e^{\prime\prime}_{ux}\rangle+b_{n,m}|e^{\prime\prime}_{ox}e^{\prime\prime}_{uy}\rangle\\
&+c_{n,m}|e^{\prime\prime}_{oy}e^{\prime\prime}_{ux}\rangle+d_{n,m}|e^{\prime\prime}_{oy}e^{\prime\prime}_{uy}\rangle]\otimes |n,m\rangle\text{,}
\end{split}
\end{equation}
where we consider the expansion within 40 oscillator quanta $(n + m)$ for the coefficient parameters.
A direct diagonalization of the total Hamiltonian with pJT and electronic part $\langle\tilde{\Phi}|\hat{H}_\text{tot}|\tilde{\Phi}\rangle$ is shown in Fig.~\ref{fig:4}(c) (see Supplementary Note 3). A converged solution demonstrates that the lowest eigenstate contains~68\% of $\tilde{A^{\prime}_{1}}$ component in the singlet manifold (and~63\% in the triplet manifold). The energy splitting between the lowest two eigenvalues are $7.1$~meV and $3.1$~meV for the singlets and triplets, respectively. Based on the degeneracy of polaronic levels, we assigned the lowest state to the $\tilde{A^{\prime}_{1}}$ and the second one to the $\tilde{E^{\prime}}$. The transition rate between polaronic states is temperature dependent, and only 0.17~ps at 100~K for singlet, as disscussed in Supplementary Note 5. Given that only the $\tilde{E^{\prime}}$ is bright, the process would require a thermal activation. Indeed, the PL intensity of UV colour centres is known to improve from low to room temperature~\cite{vokhmintsev2021temperature}, which is in line with our results. Furthermore, the position of ZPL based on the full Hamiltonian is calculated as follows
\begin{equation}
E_\text{ZPL}=E^{e}(A^{\prime}_{1})-E^{g}(A^{\prime}_{1})+\frac{1}{2}(\Lambda+\Delta)-\hbar\omega_E+\langle\tilde{\Phi}|\hat{H}_\text{tot}|\tilde{\Phi}\rangle\text{,}
\end{equation}
where $E^{e}$ and $E^{g}$ are the energies of excited state and ground state, respectively. The computed value is 4.21~eV, which is in close agreement with the experimental data. 

To further support the validity of our model calculations, we approach the $A^{\prime}_{1}$ geometry by TDDFT and CC2; however, surprisingly, we found the inconsistent results by the two electronic structure methods. More specifically, the robust CC2 approach predicts the symmetry lowering to $C_{2v}$, which is in agreement with our DFT results. This is in contrast to the TDDFT method, where the optimized structure preserves the $D_{3h}$ symmetry. In fact, this effect can be traced back to a difference between the excitation spectra in Supplementary Table 2, which can be understood as follows. Here, the energy gap between the $A^{\prime}_{1}$ and $E^{\prime}$ reflects a magnitude of the electronic coupling between the respective diabatic states. In the case of TDDFT, the value is considerably larger as compared to CC2 ($223$~meV and $178$~meV, respectively); this points to a strong coupling regime, where two diabats develop a single minima on the APES~\cite{sampaio2018kinetics}. For the 6C defect, this relaxation is particularly important, because the coupling to the $E$ phonon mode enables the intensity borrowing from the allowed $E^{\prime}$; otherwise $A^{\prime}_{1}$ state remains optically-forbidden. Another pronounced feature of TDDFT to be mentioned, is that it severely overestimates the energy gap between the lowest singlet and triplet states. Of note, the latter behaviour is largely reminiscent on the performance of this approach for the multiresonant organic emitters~\cite{pershin2019highly}. 

Having fully described the origin of the UV emission from the 6C defect, we now proceed with its spectroscopic features. For the sake of reference, we also compare our results to the experimental 4.1-eV PL signal in hBN. First, we compute the phonon sideband, which is estimated from the overlap between phonon modes in ground and excited states based on the Frank-Condon approximation~\cite{gali2009theory}. The simulated PL spectrum including the pJT distortion is shown in Fig.~\ref{fig:3}(d). Here, four prominent peaks in the phonon sideband with an averaged energy space of $180.3$~meV perfectly match the experimental PL spectrum~\cite{museur2008defect}. From these calculations, we also determine the Huang-Rhys (HR) factor, $S$, of 2.16, which is in a close agreement with the experimental results ($S = 1 - 2$). The corresponding Debye-Waller factor ($DW$), computed as $DW = e^{-S}$, is 0.11. In addition, with the CC2 approach, we obtained the HR factor of 1.3 for the hetoroatoms forming the flake. However, this value increases to 2.1, when considering the relaxation of the environment by $\Delta$SCF. 
Interestingly, as shown in Supplementary Figure 4, we also identify the low-frequency degenerate $E$-phonon modes; they represent a mutual displacement of the hBN layers and are naturally missing for the monolayer configuration. Noteworthy, at the relaxed $A^{\prime}_1$ geometry, the CC2 approach predicts that the wavefunction is governed by a single determinant with a relative contribution of 83\%. This justifies the application of the $\Delta$SCF for computing the vibronic sideband of $A^{\prime}_1$.

Next, we evaluate the radiative lifetimes based on the following expression 
\begin{equation}\label{eq:9}
\Gamma_{rad} =\frac{1}{\tau_{rad}} = \eta \frac{n_DE^3_\text{ZPL}\mu^2}{3\pi\epsilon_{0}c^3\hbar^4}\text{,}
\end{equation}
where $\epsilon_0$ is the vacuum permittivity, $\hbar$ is the reduced Planck constant, $c$ is the speed of light, $n_D = 2.5$ is the refractive index of hBN at the ZPL energy $E_\text{ZPL}$, $\mu$ is the optical transition dipole moment, and $\eta$ is the fraction of $E^{\prime}$ in the polaronic state. In $D_{3h}$ symmetry, the dipole moment operator only connects the ground state with $E^{\prime}$. Since the transition occurs within the $e^{\prime\prime}$ orbitals and the wave function overlap is large, we obtained a very short lifetime of 0.05~ns. However, the symmetry lowering makes the $\tilde{E^{\prime}}$ less bright (see Supplementray Table 3), yielding $\tau_{rad}$ = 1.54~ns at room temperature (2~ns at 150~K for the SPE experiment~\cite{bourrellier2016bright}). This value is temperature-dependent considering the thermal occupation of $\tilde{E^{\prime}}$. Nonetheless, it is very close to the observed $\sim$1.1~ns~\cite{museur2008defect}.

Beside the radiative decay, we have also explored a possibility of the non-radiative transition to the triplet manifold through the intersystem crossing (ISC). This process is mainly governed by the spin-orbit coupling (SOC), and the possible pathways are depicted in Fig.~\ref{fig:4}(d). The SOC interaction can be expressed as
\begin{equation}\label{eq:soc}
\hat{H}_\text{SO} = \sum_{k}\lambda_{x,y}(l^x_ks^x_k+l^y_ks^y_k)+\lambda_{z}(l^z_ks^z_k+l^z_ks^z_k)\text{,}
\end{equation}
where  $\lambda_{x,y}$ are the non-axial components, while  $\lambda_{z}$ is an axial component. In particular, $\lambda_{x,y}$ couples the triplet states with the non-zero spin projections ($m_s = \pm 1$) with singlets of different electronic configuration. In turn, $\lambda_{z}$ links states with $m_s = 0$ spin projections with the states of the same electronic configuration. Since all the excited states in our system have the same electronic configuration $|e^{\prime\prime}e^{\prime\prime}\rangle$ only the axial part is non-vanishing. The SOC splits $^3E^{\prime}$  into $m_s = \pm 1$ sub-states $A_{1,2}$ and $E_{1,2}$ with the $m_s = 0$ $E_{x,y}$ state. In addition, $^1A^{\prime}_1$ could also decay to $^3E^{\prime}$ due to the mixture with $^1E^{\prime}$. The ISC rate from singlet to triplet can be calculated by~\cite{goldman2015state}
\begin{equation}\label{eq:nonradiative}
\tau_\text{ISC} = 4\pi\hbar{\lambda^2_z}F(\Delta E)\text{,}
\end{equation}
where $F$ is the spectral function of vibrational overlap between the singlets and triplet states, and $\Delta E$ is the energy splitting between singlet and triplet levels. From the TDDFT calculations, we found the largest value of $\lambda_{z}$ of only $1.5$~GHz. Given a considerable energy gap between the states, this translates into the enormously large $\tau_\text{ISC}$, and disables the ISC process in the zeroth-order (see Supplementary Figure 2). Yet, we note that the triplet manifold may be populated via a nongeminate recombination of hot charge carriers achieved by a two-photon absorption process.

Another non-radiative transition occurs between the $A^{\prime}_1$ and the lower-lying $A^{\prime}_2$ in singlet manifold. This process could bleach the fluorescence if it is faster than the radiative lifetime. We evaluate the transition rate by calculating the electron-phonon coupling between the $e$ orbitals as discussed in Supplementary Note 4. In a low temperature limit, the computed rate is $509$~MHz ($1.98$~ns) which is slower than the above mentioned radiative rate. The optimal quantum efficiency for the defect to~52\% at 300~K, however, this is influenced by temperature which can change the distribution between the dark and bright polaronic states. As discussed in Supplementary Note 5, the brightness increases as the temperature is elevated.
We note that the non-radiative decay via phonons from the singlet $A^{\prime}_2$ towards the ground state is very slow due to the large gap between the two, thus recombination of hot charge carriers via two-photon absorption process is the likely process to get to the ground state once the electron is scattered to the dark singlet $A^{\prime}_2$ state.

Finally, after identifying the 6C defect as a promising candidate for the UV emission, we compare its properties with those of 4C and \cncb. While the \cncb\ defect was described elsewhere~\cite{mackoit2019carbon}, for 4C$_\text{pair}$ we computed the ZPL of $\sim$4.4~eV and the HR factor of $1.9$. As demonstrated in Fig.~\ref{fig:4}(f) all three defects exhibit a remarkably similar sideband, while \cncb\ shows a slightly smaller energy space between the phonon replicas due to a smaller HR factor of 1.6. The minor differences between those are seen in the intensities of the replicas at the lower energies. These findings are in line with a recent experimental work, where a continuous distribution of ZPL lines around 4.1 eV ~\cite{pelini2019shallow} is observed. Therefore, other means than a simple PL characterisation would be of help to distinguish between the actual carbon configurations. In particular, to confirm the involvement of carbon in a colour centre, it is commonplace to use the isotopic purification method, that incorporates ${}^{13}$C into the lattice of the material during its growth. Here, we determine the isotopic shift in the emission energy and sideband for the 6C defect associated by replacing ${}^{12}$C with ${}^{13}$C isotope. First, we calculated the sideband with 100\% of ${}^{13}$C isotopes and found that the HR factor reduces to 1.78, see Fig.~\ref{fig:4}(e). In turn, the phonon replicas show a blue shift by $\sim$10~meV. For comparison, the isotopic effect on \cncb\ introduces and the 4C$_\text{pair}$ has similar blue shift but with different values as shown in Supplementary Figure 4. This might provide a feasible way to differentiate the configurations for emissions.

%
\section{Summary and conclusion}
\label{sec:summary}

In summary, based on an extensive theoretical investigation, we explored the potential of substitution carbon defects to develop UV single-photon sources in hBN. By conducting a systematic study on seventeen defect configurations, we found that carbon atoms are preferentially arranged into chains, which are stabilized to a formation of the energetically favourable C-C bonds. Of those defect configurations, we identified several potential candidates for the UV emission, including \cncb, 4C, and 6C defects, since they feature a photo-stable (neutral) charge state. Furthermore, we specifically focused on the electronic and optical properties of the 6C defect of which configuraton was observed in experiments. We found that it exhibits a highly non-trivial emission mechanism where the second excited state is optically activated by the product Jahn-Teller effect. More specifically, the ZPL is computed at 4.21 eV and the HR factor is found to be 2.1; the simulated PL spectrum shows the phonon replicas with an energy spacing of 180~meV; the upper limit of estimated radiative lifetime is ,$\sim$1.17~ns. All these properties closely resemble the PL signal that is natively present in many hBN samples. Given the relative low formation energy and complete agreement with the experimental measurements, these results outline the 6C defect as a plausible source of the observed UV emission. However, by comparing the properties of 6C with the other chain defects, we found the remarkable similarities in the positions of the ZPL and vibronic sideband. Based on this data, we infer that the 4.1-eV PL signal likely appears as a commutative effect from different types of point defects. We show that the colour centres can be distinguished by the respective isotope shift of their sideband. The 6C stands out from these C defects in two ways: first, it exhibits a temeperature dependent brightness; second, it shows a substantial change of the phonon properties from monolayer to multilayer configurations. We anticipate that the latter can represent as a useful metric to characterize the number of layers in hBN by the simple spectroscopic measurements. Furthermore, it is likely that 6C ring defect is responsible for the temperature dependency of the 4.1-eV emission~\cite{vokhmintsev2021temperature} from the family of 4.1-eV emitters.

\section{Methods} 
\label{sec:methods}

\subsection{Details on DFT calculations}
The calculations were performed based on the spin-polarized DFT within the Kohn-Sham scheme as implemented in Vienna
\textit{ab initio} simulation package (VASP)~\cite{kresse1996efficiency, kresse1996efficient}. A standard projector augmented wave (PAW) formalism~\cite{blochl1994projector, kresse1999ultrasoft} is applied to accurately describe the spin density of valence electrons close to nuclei. The carbon defects were embedded in a $7\times7$ bulk supercell with 196 atoms. The atoms were fully relaxed with a plane wave cutoff energy of 450~eV until the forces acting on ions were less than 0.01~eV/\AA. The Brillouin-zone was sampled by the single $\Gamma$-point scheme. The screened hybrid density functional of Heyd, Scuseria, and Ernzerhof (HSE)~\cite{heyd2003hybrid} was used to optimize the structure and calculate the electronic properties. By changing the $\alpha$ parameter, we modified a part of nonlocal Hartree-Fock exchange to the generalized gradient approximation of Perdew, Burke, and Ernzerhof (PBE)~\cite{perdew1996generalized} with fraction $\alpha$ to adjust the calculated band gap. Here, $\alpha$ = 0.32 was used which could reproduce the experimental band gap about 6~eV. The optimized interlayer distance was 3.37~\AA\ with DFT-D3 method of Grimme~\cite{grimme2010consistent}. The excited states were calculated by $\Delta$SCF method~\cite{gali2009theory}.
The defect formation energies $E_f$ was calculated according to the following equation,
\begin{equation}\label{eq:formation}
\begin{split}
E^q_f = &E^q_d - E_\text{p} - n_\text{C}\mu_\text{C} + n_\text{B}\mu_\text{B} + n_\text{N}\mu_\text{N} + \\ 
&q\left(\epsilon^\text{p}_\text{VBM} + \epsilon_\text{Fermi}\right) + E_\text{corr}\left(q\right)\text{,}
\end{split}
\end{equation}
where $E_d^q$ is the total energy of hBN model with defect at $q$ charge state and $E_\text{p}$ is the total energy of hBN layer without defect. $\mu_\text{C}$ is the chemical potential of carbon and can be derived from pure graphite.  For N-rich condition, the chemical potential $\mu_\text{N}$ = $1/2E(\text{N}_2)$, which is half of nitrogen gas molecule. For N-poor condition, the chemical potential $\mu_\text{B}$ is derived from pure bulk boron and $\mu_\text{BN}$ = $\mu_\text{B}$ + $\mu_\text{N}$. The Fermi level $\epsilon_\text{Fermi}$ represents the chemical potential of electron reservoir and it is aligned to the valence band maximum (VBM) energy of perfect hBN, $\epsilon^\text{p}_\text{VBM}$. The $E_\text{corr}\left(q\right)$ is the correction term for the charged system due to the existence of electrostatic interactions with periodic condition. The charge correction terms were computed by SXDEFECTALIGN code from Freysoldt method~\cite{freysoldt2018first}.
\subsection{Post-Hartree-Fock methods and TDDFT calculations}
For the excited state calculations, the 6C defect was incorporated into a flake of hBN, containing 27 boron and 27 nitrogen atoms. The dangling bonds were passivated by hydrogen atoms. The calculations with the second-order approximate coupled cluster singles and doubles model (CC2)~\cite{christiansen1995second} and the algebraic diagrammatic construction method, ADC(2)~\cite{schirmer1982beyond} were performed with Turbomole code~\cite{ahlrichs1989electronic, TURBOMOLE}.

The results of time-dependent (TD) DFT and n-electron valence state perturbation theory, NEVPT2(4,4)~\cite{angeli2001introduction} were obtained by ORCA code~\cite{neese2018software}. In all calculations, we used cc-pVDZ basis set~\cite{dunning1989gaussian} and considered the PBE0 density functional~\cite{perdew1996rationale} for TDDFT. To compute the HR factor by CC2, we first relaxed the system in both ground and excited states. Then, all atoms but hydrogen atoms were incorporated into the periodic lattice of hBN preserving their equilibrium positions. In this way, we could reach a consistent description of the sideband with the $\Delta$SCF method, thereby relying on the PBE normal modes in both cases.

\section*{Author contribution}
A.\ G.\ and S.\ L.\ conceived the work. S.\ L.\ carried out the DFT calculations and related analysis. A.\ P.\ performed the post-Hartree fock and TDDFT calculations. S.\ L.\ and A.\ P.\ wrote the manuscript. All authors discussed the results and contributed to the improvement of the manuscript. A.\ G.\ supervised the entire project.

\section*{Competing interests}
The authors declare that there are no competing interests.

\section*{Data Availability}
The data that support the findings of this study are available from the corresponding author upon reasonable request.

%
%
\begin{acknowledgments}
AG acknowledges the Hungarian NKFIH grant No.~KKP129866 of the 
National Excellence Program of Quantum-coherent materials project, and the support for the Quantum Information National Laboratory from the Ministry of Innovation and Technology of Hungary, and the EU H2020 Quantum Technology Flagship project ASTERIQS (Grant No.\ 820394). We acknowledge that the results of this research have been achieved using the DECI resource Eagle based in Poland at Poznan with support from the PRACE aisbl. During the paper submission, a paper about thermodynamics of carbon point defects in hBN by Maciaszek \textit{et al.} has appeared on arXiv~\cite{maciaszek2021thermodynamics}.
\end{acknowledgments}

%
%
\bibliography{mainref}

\end{document}